\def\year{2020}\relax
\documentclass[letterpaper]{article} % DO NOT CHANGE THIS
\usepackage{aaai20}  % DO NOT CHANGE THIS
\usepackage{times}  % DO NOT CHANGE THIS
\usepackage{helvet} % DO NOT CHANGE THIS
\usepackage{courier}  % DO NOT CHANGE THIS
\usepackage[hyphens]{url}  % DO NOT CHANGE THIS
\usepackage{graphicx} % DO NOT CHANGE THIS
\usepackage{dblfloatfix}
\usepackage{hyperref}
\usepackage{listings}
\usepackage[utf8]{inputenc}
\usepackage{graphicx}
\usepackage{tabularx}
\usepackage{multirow}
\usepackage{multicol}
\usepackage{rotating}
\usepackage{xcolor}
\usepackage{forest}
\usepackage{amsmath}
\usepackage{booktabs}
\usepackage{dcolumn}
\newcolumntype{d}[1]{D{.}{.}{#1}}
\usepackage{siunitx}
\usepackage{amsmath}

\usepackage{graphicx}  % DO NOT CHANGE THIS
\title{Quasi-experimental Designs for Assessing\\ Response on Social Media to Policy Changes}
\author{Yijun Tian, Rumi Chunara\\ \Large%\textbf{AAAI Style Contributions by
New York University\\ %If you have multiple authors and multiple affiliations
eddie.tian@nyu.edu, rumi.chunara@nyu.edu % email address must be in roman text type, not monospace or sans serif
}
 
\begin{document}

\maketitle

\begin{abstract}
Regulation of tobacco products is rapidly evolving. Understanding public sentiment in response to changes is very important as authorities assess how to effectively protect population health. Social media systems are widely recognized to be useful for collecting data about human preferences and perceptions. However, how social media data may be used, in rapid policy change settings, given challenges of narrow time periods and specific locations and non-representative the population using social media is an open question. In this paper we apply quasi-experimental designs, which have been used previously in observational data such as social media, to control for time and location confounders on social media, and then use content analysis of Twitter and Reddit posts to illustrate the content of reactions to tobacco flavor bans and the effect of taxation on e-cigarettes. Conclusions distill the potential role of social media in settings of rapidly changing regulation, in complement to what is learned by traditional denominator-based representative surveys.
\end{abstract}

\section{Introduction}
%\ ecig/tobacco/changes
Tobacco continues to be a global public health threat,  killing more than five million people each year \cite{world2010technical}. While the harm from combustible tobacco is well known, tobacco products continue to be widely and legally available. Moreover, the landscape of tobacco products is rapidly evolving, and to minimize  potential public health harm, the US Food and Drug Administration, as well as state and local agencies engage in regulatory actions to discourage use \cite{hefler2018changing}. In fact, the week of initially writing this paper, the CDC, the U.S. Food and Drug Administration (FDA), state and local health departments, and other clinical and public health partners began investigating a multi-state outbreak of lung disease associated with electronic cigarette (``e-cigarette'') use, and many cases and deaths in states that \emph{already have e-cigarette taxes}. This outbreak, alongside continued heavy use of tobacco products highlights that a continuous understanding of public sentiment on tobacco products and regulation efforts is necessary \cite{cdctobacco}. Such proactive efforts can serve to assess the impact of regulations on public sentiment; to understand if they may be discouraging use or changing preferences, especially in youth who are significant consumers of new tobacco products such as e-cigarettes \cite{cdctobacco}. 

Taxation is a regulatory tool that has been used over many years for tobacco products. While there is expert-backed evidence that on the effectiveness of increased tobacco  taxes and prices in reducing overall tobacco consumption, prevalence of tobacco use and improvement of public health, it is not an all-encompassing solution. Indeed, different types of tobacco users may be less sensitive to changes in prices, or individuals may start to lean on other products if regulations are incurred \cite{hefler2018changing}. Moreover, evidence on regulation around new tobacco products like e-cigarettes is limited. E-cigarettes have had a quickly evolving population penetrance alongside unexpected findings and changes in their use since their initial availability on the market in 2003. Originally, e-cigarettes have been viewed in some ways, as an important pathway for those trying to quit combustible tobacco products (e.g. cigarettes, cigars, etc.). However in a short recent past, the surge of young users has resulted in concerning health behaviors and outcomes. In particular, increased use in youth alongside introduction of flavors and promotion by social media influencers has created new phenomena by which e-cigarette use has become very prevalent \cite{gostin2014cigarettes}. In sum, these rapid changes to the landscape have accelerated the need for regulation, which has happened and is going to happen at multiple spatial scales in the near future \cite{taxfound}. 
s
Policy changes are typically assessed in retrospect through surveys. Though surveys are robust and can be targeted to specific populations, there are severe time-lag, recall and information bias issues; individuals may not feel comfortable to disclose risky behaviors in surveys \cite{chunara2017denominator}. Also, policy changes can take effect at multiple geographic levels; cities, states, nation-wide, and can impact multiple products, making gathering information on regulatory effects complex and hard to measure via surveys. This context incurs the need for rapid, easy to measure effect of policies on public sentiment on multiple products, and in multiple places, in a timely manner. 

%the evidence examined is mixed and varies with a country's circumstances; for example, in settings where there is ready access to low or untaxed and inexpensive tobacco products, low income tobacco users may be less sensitive to changes in prices [cite]. 

%including by preventing initiation and uptake among young people, promoting cessation among current users, and lowering consumption among those who continue to use.

%\ policies + use of social media + ecig on social media
Social media systems are widely recognized, especially in health, to be useful for collecting data about human preferences, sentiments, and reactions \cite{chunara2013assessing,de2013predicting,bai2015social,wang2018s}. While there are limitations in understanding the nature of the population represented on social media, the volume of data and known high use in young populations, makes it a informative source to especially learn about immediate effects of changes, in relation to other products and places. In particular, adolescents and young adults are among the groups most heavily engaged in social media including with regard to posting about their health and social behaviors \cite{salimian2014averting}. Products such as e-cigarettes are discussed heavily on social media; this discussion has been used to better understand e-cigarette attitudes and behaviors by public health and communication professionals \cite{cole2015social}. This synergy of high social media in use and concerning use of tobacco products in youth, make it a relevant medium through which to assess the pulse on specific topics. 

However, there are major concerns with drawing population-level conclusions from social media across time or locations because of: the opt-in nature of contributing data (affecting types of populations represented), temporal confounders, and differences between places that are not explicitly measured \cite{chunara2017denominator,tufekci2014engineering,olteanu2019social}. Moreover, when zooming into a specific context (time and place), concerns can be elevated due to further narrowing of the population. Quasi-experimental designs (QEDs) have been used to address methodological challenges of drawing conclusions from observational data sources like social media data. Commonly used in social sciences to discover causal knowledge from observational data, QEDs can be used to discover knowledge from social media data by decreasing confounding \cite{oktay2010causal}. Given the rapidly changing tobacco and regulation landscape, public reception and response to regulations are of significant importance. Understanding public conversations allows public health and communication professionals to identify trends in attitudes and behaviors and to adjust course on regulations already in effect. Towards the need to understand public sentiment on tobacco policy changes at multiple scales and across multiple types of regulations, we conduct the first study of policy response on social media in relation to recent tobacco regulations in the United States. We use data from social media to investigate two research questions. We use rigorous quasi-experimental study designs to decrease confounding and draw conclusions as best as possible, followed with a content analysis from Reddit to expound on and validate findings regarding the content from Twitter. Specific research questions addressed are:\\
\textbf{RQ1.} Is there an effect on online sentiment in San Francisco for different tobacco products based on different stages of a tobacco flavor ban?\\
\textbf{RQ2.} Upon implementation of a state-level e-cigarette tax, is there an effect on online e-cigarette sentiment in other states?\\

%The prevalence of e-cigarette use has risen rapidly since introduction of this product to the United States in 2007.

\section{Related Work}
%\ policies + use of social media
%\ ecig/tobacco/changes
\subsection{Tobacco Discussion on Social Media} Tobacco is a well-discussed topic on social media. Social media is particularly relevant given the synergy of demographics of Twitter users and of those increasingly using new tobacco products. Most studies have focused on qualitative perception, sentiment and topic analyses, and identifying relevant hashtags \cite{lazard2016cigarette}. Studies of tobacco-related topics show that a large majority of posts on tobacco tend to center on experience sharing, with minor contributions in terms of promotion or specific to regulation debates \cite{huang2017high,krauss2015peer}. We could not find any studies analyzing social media sentiment about tobacco policy implementation. 

\subsection{Social Media and Policy}
As public health interventions are often at the population-scale, monitoring public opinion is an important activity. Public response to substance use regulations are of particular interest. For example, public opinion has been one factor affecting change in policies designed to reduce underage alcohol use \cite{latimer2003measuring}. In general, the degree to which the public supports or opposes the policy can help with designing policies amenable to compliance or understanding other public needs \cite{latimer2003measuring}. These public measures are typically sourced through surveys, which have been criticized for poor measure reliability and validity \cite{chunara2017denominator}. 
%Along with topics on specific products such as swishers, Juul (e cigarette brand)

Social media has been used to examine public discourse around ideological issues that also overlap with policy, for example the abortion debate \cite{sharma2017analyzing,zhang2016gender}. In the work by Sharma et al., authors specifically analyze textual and psycholinguistic cues in different ideological categories on this topic; in contrast to looking at specific temporal changes. This work also highlights how social media can be used as a platform to express views on contentious issues (such as tobacco policy changes as highlighted), making its analysis on such topics particularly of interest. In Zhang and Counts, people’s expressions of opinion on abortion in relation to the policy changes are studied. Both papers find discussion relating to ideological discourse on abortion. Work by Saha et al. leverages social media as a passive sensor of stress to quantify and examine stress responses \textit{after} gun violence on college campuses \cite{saha2017modeling}. This work uses data from Reddit, which offers a large amount of location-specific text content (e.g. from campus sub-reddits). Stress and abortion are also arguably more common than discussion specific to tobacco/e-cigarettes. For analysis of posts over time, this work used an approach based on both difference in means and linear regression. Although, the recommendations this paper examines effects of, are also made online. In regards to policy response, some work in the social media literature has examined the nature of and engagement around content shared in \textit{online} campaigns such as for mental health  \cite{saha2019computational}. 

Results of such work suggest social media can play an important role role in understanding of public sentiment towards offline regulation. Each topic also reflects the specific way social media can complement survey-based opinion sampling; to provide an unobtrusive window into sensitive topics such as abortion or stress. However, none have focused on looking immediate response to an offline policy change, or that response in different spatial locations.

\subsection{Quasi-experimental Designs} 
Quasi-experimental designs (QEDs) are one approach that have been used to decrease confounding in observational data; they a type of design that is often used in circumstances when random assignment of treatment is either impossible or infeasible. QEDs have a general idea similar to that of a randomized experiments; they work by identifying an experimental unit that has undergone treatment and comparing it to another experimental unit that has not
undergone treatment but which is similar in other aspects \cite{shadish2002experimental}. Such  analytical  designs  have  previously been used with social media data \cite{oktay2010causal}. It should be noted that by comparing to units with similar properties, these designs can decrease confounding by unobserved variables (e.g. over time or by place), which are \textit{some} of the threats to validity when drawing conclusions from social media data \cite{chunara2017denominator,olteanu2019social}. Such  analytical  designs  have  previously been used with social media data \cite{oktay2010causal}. This paper applied two different QEDs to demonstrate how one can overcome some threats to validity in social media data through constructing careful study designs. Amongst these methods were an interrupted time series design to mitigate temporal confounders and a natural experiment design to assess effect of one single (exogenous) change. Before implementation of each method, we describe the tobacco regulation, relevance of the method and what kind of concerns it mitigates.

\section{Experiment One: Tobacco Flavor Ban}
\subsection{Background on San Francisco Ban}
The San Francisco Health Code, Article 19Q:  Prohibiting the Sales of Flavored Tobacco Products was approved by San Francisco voters on June 5, 2018. The flavor ban encompassed all forms of flavored tobacco. Indeed, some of the most critical products and flavors that are thought to be most appealing to youth are in e-cigarettes. Thus we considered social media data regarding i) all tobacco and ii) e-cigarettes in San Francisco for this analysis.

Understanding public sentiment on flavored tobacco is important to examine in relation to implementation of such a ban due to a variety of opposing concerns on the topic including: anticipation of harm to small business economies, lobbying by tobacco companies who see vaping as the future of their business, pushback regarding government overreach on regulating behaviors and choices and public health indication that flavors lead to addiction and health concerns \cite{hefler2018changing}. Assessing this sentiment is particularly important to do in the short-term, as multiple other municipalities are rapidly following suite (e.g. in nearby Oakland). Monitoring sentiment in relation to a ban is important over time before and after the ban as well; even though a ban may take effect on a certain date, there is time needed for businesses to implement the policy as well as anticipatory action that individuals may take (increase purchases, etc.) in advance of a ban \cite{bernal2017interrupted}. The San Francisco policy can be divided into four time intervals: proposal (June 20, 2017), approval (June 5, 2018) and enforcement (Jan 1, 2019).

%therefore we use interrupted time series approach in order to assess this over time, and in a controlled way (control for underlying tweet trends over time, etc.)

% plots of SF ecig tweets by month - from github

% compute statistic significance of difference in slopes throughout policy change (RC to look up statistic to use)

% flavor ecig&tobacco tweets qualitative look

% YT to run pre-trained tobacco classifier, sentiment analysis.

%\background on policy change details

%\sentiment analysis
% why The Interrupted Time-Series Design
\subsection{Data Collection and Processing}
\paragraph{Tobacco Classification}
Data was obtained from the Twitter public API. We limited our search to Tweets with either point or polygon geo-location information, as location information was critical for our study. Our data access was limited to April 2016 through April 2019. Therefore, data for this experiment, the first time interval prior to the proposal consists of 14 months (April 2016 - May 2017), for the second time period between proposal and approval consists of 12 months (June 2017 - May 2018), the time period between approval and enforcement is 7 months (June 2018 - December 2018), and there are 4 months remaining between the enforcement until the end of the dataset (January 2019 - April 2019). For classifying Tweets that discuss Tobacco use, we used a previously developed Tobacco classifier [citation blinded]. This approach uses a three-stage classification to ensure the tweet relates to a) tobacco, b) a person discussing their own use and c) at that time. Tobacco use in relation to the flavor ban is also of interest in this study, so this is appropriate. A grid search results in logistic regression being used for each classifier. Regularization coefficient was C = 0.01. F1 scores for the three classifier stages were 0.86, 0.98, 0.75. Accuracy for each stage was 0.96, 0.77 and 0.77. More details regarding training, testing and evaluation of the classifier are previously published; we used the exact same three-stage classification process to resolve tobacco Tweets here.

\paragraph{E-cigarette Tweet Parsing}
%\classification vs keyword
Given the high amount of interest in flavors for e-cigarettes, understanding e-cigarette use and discussion in relation to the flavor ban is of particular interest. We first developed a classifier for e-cigarettes, with the aim of using it to find all related e-cigarette Tweets. In sum, we found that training a classifier on labelled Tweets was no better than using a keyword list to filter; there were no other combinations of words that were indicative of e-cigarette Tweets. Details for the process follow, along with resulting findings and description of how we created the e-cigarette related keyword list.

For creating labelled data to train the classifier, we first developed a detailed keyword list to filter out e-cigarette relevant Tweets from those obtained via the Twitter API, for use in training a classifier to identify e-cigarette Tweets comprehensively. The keyword list was derived from: 1) findings from papers that have examined e-cigarette content on social media \cite{allem2017cigarette,cole2015social,lazard2016cigarette,aphinyanaphongs2016text}, keywords regarding e-cigarette policies (codes of the relevant senate bills and their linguistic variants, e.g. ``sb 140'' and ``sb140'' for sb140 and ``sb 24'' and ``sb24'' for sb24), popular e-cigarette hashtags related to cessation (`NotAReplacement' and `tobaccofreekids') which we ascsertained from manually reviewing e-cigarette Tweets, and keywords from reviewing the Twitter accounts of e-cigarette brands (which ended up being a list of e-cigarette brand names, e.g. Juul, Blu, VaporFi, etc.). All keywords are listed in the Appendix. 

%Necessary spaces are included before and after some keyword. 
%Combining these with a set of Tweets with no 
We then randomly selected 800 Tweets, and filtered them by the e-cigarette keywords for labelling. We used services of Figure Eight, acquiring two labels for each Tweet. To ensure quality of the annotation process, we used 41 test questions and each contributor was required to label at least 10 test questions. Contributors with agreement under 70\% on the test questions were considered untrusted, a common threshold when using Figure Eight \cite{aggarwal2019classification}, and their judgements were removed from our task. To further improve annotation quality, the task was very clearly specified in Figure Eight, with several examples around edge cases that we ascertained using early tests of labeling. In sum, we asked labelers to decide if the Tweet had any mention of anything related to e-cigarettes (for example description of past, present or future use of an e-cigarette, sharing any attitude or feeling about e-cigarettes or talking about any e-cigarette brand or any e-cigarette related device (including any pods, accessories, etc). Agreement for labeling was 91.4\%. A member of our team resolved conflicts for any Tweets with differing labels from trusted contributors. Examining results of the labelling procedure showed that mis-classifications were generally due to misunderstanding about e-cigarettes; difficult to resolve even with instructions and examples. For example, false negatives were generally due to unfamiliarity with e-cigarette synonyms and e-cigarette brand names; false positives could be attributed to confusion about e-cigarette keywords and word contained in e-cigarette keywords (e.g. treating `vapormax', `vaporrub, `vaporwave' as e-cigarette-related), or if a keyword was included in the account name. Through examining true negatives we further refined keywords in the list (for example we found that the word `krave' which is an e-cigarette brand, often appears in Tweets in other contexts such as a synonym for crave), so we removed this from the keyword list. 

Examining results of the labeling, true positive and negatives, we found that by using the updated keyword list, there is a very low probability of false positives (we found zero in the set of Tweets labelled and no erroneous categorization of any other Tweets either). Thus, Tweets captured through this list would most likely be e-cigarette related. In any case, given that there may be combinations of words and phrases that could still be indicative of e-cigarette Tweets, we then built a classifier, to identify such potential features and find Tweets related to e-cigarettes that would not be found based on the keyword list. We used a logistic regression classifier, and a training data set composed of a balanced training data set composed of e-cigarette positive Tweets from the labelled data and Tweets without any of these keywords. However, through systematic analysis of the input data balance and classifier threshold, we could not find evidence of any e-cigarette Tweets that would not contain the sourced keywords. Therefore, we concluded that using the keyword list to source Tweets would be sufficient. There is a possibility that we would miss some e-cigarette related Tweets by using the keyword list only, though given that these would be very difficult to identify, we trade off good precision (a classifier would also return false positives), for slightly lower recall.

\paragraph{Flavored Tobacco Tweet Parsing}
As the San Francisco ban was on flavored tobacco specifically, we also sought to identify flavored-tobacco related Tweets. To do this, we filtered tobacco-related Tweets (identified using the classifier described above) by a list of flavor keywords. This list (included in the Appendix) was developed using product lists from tobacco products (most of which were e-cigarettes). %Our team manually examined resulting Tweets to assess if they were in fact flavor-tobacco related and found this to be the case with high probability. Most of the ``flavored tobacco'' Tweets included the word ``flavor'', however other flavors were described, albeit in much smaller proportions.

\paragraph{Source of Tweets}
Given increasing recognition of the possibility of automated accounts on Twitter, especially those sourcing misinformation, and in relation to health topics \cite{broniatowski2018weaponized}, we assessed the possibility of bots accounts in the resulting tobacco and e-cigarette Tweets data. To do so, we used seven available lists of bot accounts (the same ones used in multiple other social media studies \cite{broniatowski2018weaponized,relia2019race}) to identify bot accounts. Among e-cigarette tweets, tobacco tweets, and flavored tobacco Tweets we collected, none of them were composed by users who appear in the bot accounts list. As another qualitative check, we also calculated the average number of days all of the tobacco, e-cigarette user accounts have existed and the average number of Tweets they posted each day. We did not observe any extreme outliers. Therefore, it is unlikely that there were any bot accounts in our dataset. It should also be noted that a very low proportion of the filtered e-cigarette Tweets were commercial in nature. We extracted 1000 Tweets randomly from all San Francisco e-cigarette Tweets over three years. We manually examined these Tweets, and found that 52 out of the 1000 were from a commercial source (opposed to an individual).

\begin{table}[b!]
\centering
\caption{Experiment 1: Summary of $F_{tweets}$: avg number of Tweets per month in each implementation phase, $F_{users}$: avg number of users per month posting related Tweets, $F_{followers}$ and $F_{friends}$: (mean, median) for each user during each phase of policy implementation.}

\label{table:summary_table_exp_1}
\resizebox{\columnwidth}{!}{
\begin{tabular}{|c|c|c|c|c|c|}
% if space available
% \addlinespace[1ex]
\hline
$Tweets$ & $Time Period$ & $F_{tweets}$ & $F_{users}$ & $F_{followers}$ & $F_{friends}$  \\
\hline
\hline

\multirow{4}{*}{E-Cigarette} & Before Proposal & 105 & 78 & 7949, 480 & 1547, 447  \\
\cline{2-6}
 & Proposal - Approval & 86 & 65 & 9145, 530 & 856, 487 \\
\cline{2-6}
 & Approval - Enforcement & 107 & 85 & 19412, 508 & 1166, 466  \\
\cline{2-6}
 & After Enforcement & 100 & 80 & 8791, 685 & 1310, 557  \\
\hline
\hline

\multirow{4}{*}{Tobacco} & Before Proposal & 746 & 581 & 5698, 472 & 1220, 420  \\
\cline{2-6}
 & Proposal - Approval & 588 & 454 & 3971, 476 & 978, 449  \\
\cline{2-6}
 & Approval - Enforcement & 597 & 479 & 8027, 505 & 985, 486  \\
\cline{2-6}
 & After Enforcement & 561 & 418 & 5678, 510 & 1128, 519 \\
\hline
\hline

\multirow{4}{*}{Flavored Tobacco} & Before Proposal & 65 & 61 & 2706, 469 & 957, 444  \\
\cline{2-6}
 & Proposal - Approval & 40 & 38 & 3117, 468 & 1131, 439  \\
\cline{2-6}
 & Approval - Enforcement & 44 & 41 & 1817, 504 & 1053, 438 \\
\cline{2-6}
 & After Enforcement & 42 & 38 & 3740, 548 & 859, 557  \\
\hline
\end{tabular}
}
\end{table}

%\paragraph{Explaination for Table \ref{table:summary_table_exp_1}}

\subsection{Sentiment Trends}
First, to understand if there were differing sentiment trends and changes during each time period, we examined the trend in positive and negative sentiment tobacco and e-cigarette Tweets in the four different intervals; preceding and following the proposal, approval and enforcement of the ban. All Tweets (tobacco, e-cigarette and flavored tobacco) were categorized into positive, negative or neutral sentiment using Vader. Vader is a rule-based model for sentiment analysis of social media text \cite{hutto2014vader}. To then examine sentiment trends over time, we fit linear regression models in each time period (for each of the tobacco, e-cigarette and flavored tobacco Tweets in San Francisco). We added an interaction term to each model to allow us to test the hypothesis that the relationship between the trend in proportion of e-cigarette/tobacco/flavored tobacco Tweets was different for positive versus negative sentiment. %This approach has the advantage of not requiring that the data come from independent normal random variables (for example if we used a t-test to assess statistical significance of difference in slopes).

\paragraph{Interrupted Time Series Analysis}
Next, to understand if the trend in Tweets and positive/negative sentiment for each of tobacco/e-cigarette and flavored tobacco Tweets significantly changed during time periods related to the ban, we used an interrupted time series design. 
In the interrupted time series design, we observe an outcome variable for a certain time interval, $\Delta t$, before a treatment and after the treatment. Each segment of the series is allowed to exhibit both a level and a trend. These are both observed before and after a trend that follows an intervention. A change in level, e.g. a jump or drop in the outcome after the intervention, constitutes an abrupt intervention effect. The main advantage to this approach is that the observation over $\Delta t$ is by accounting for trends in the same time series over time, even without a control group, we can control for any intrinsic changes within the time-series and therefore rule out temporal threats to validity (e.g. from unmeasured confounding variables). In contrast to cross-sectional observational studies, segmented regression analysis of interrupted time series data allows analysts to control for prior trends in the outcome and to study the dynamics of change in response to an intervention. In other words, it would be insufficient to simply say that e-cigarette Tweets increased at a certain time, without comparing the trend to that before/after an event of interest (e.g. accounting for situations such as overall Tweeting levels changing over time for any other reason). The only main remaining issue is if the outcome (number of Tweets) changes for an unrelated reason at the exact same time intervals being examined, which is very unlikely.

For computing differences before and after policy-related events, we used a two-sample t-test of the mean proportion of Tweets by month for the before and after intervals to test the hypothesis that the difference between before and after is equal to zero. A t-statistic is often used, in preference to a z-score for example, when the sample size is small.

%Experiment 1:
%significant increase in tweeting after proposal (intercept jump)
%negative tweets increasing after proposal, and approval positive decreasing

%  exp1 proportion Figure
\begin{figure*}[ht!]
\caption{E-cigarette, tobacco and flavored tobacco Tweet trends in San Francisco. 95\% confidence intervals shaded.}
\includegraphics[width=0.94\textwidth]{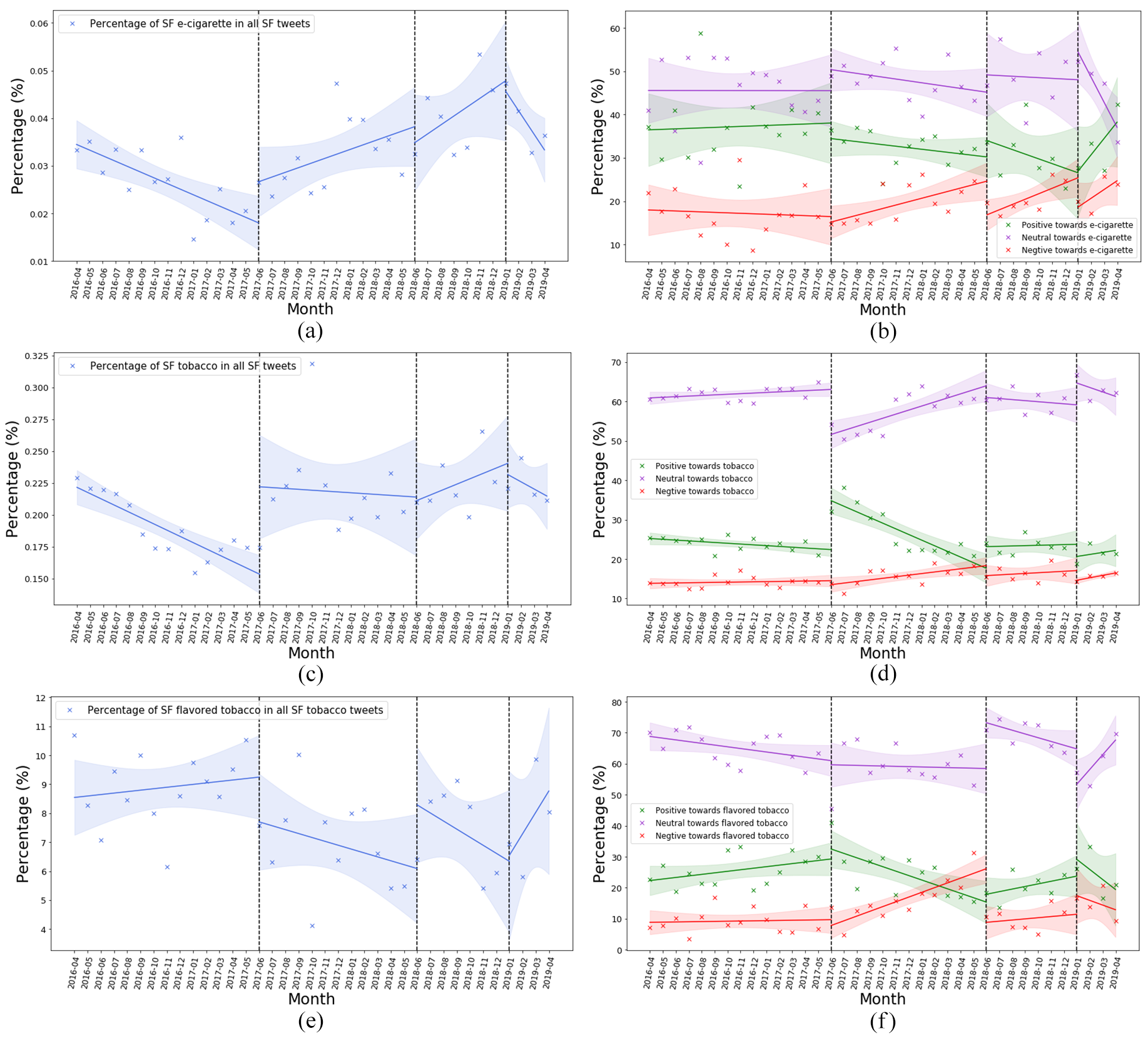}
\centering
\label{fig:exp1_ecig_tobacco}
\end{figure*}

\begin{table}[h!]
\centering
\caption{Slope and intercept values of e-cigarette, tobacco and flavored tobacco Tweet trends in San Francisco, by sentiment, and significance in change with respect to proposal, approval and enforcement of flavored tobacco ban events.}
\label{table:t-test}
\resizebox{0.97\columnwidth}{!}{
\begin{tabular}{|c|c|c|l|l|}
% if space available
% \addlinespace[1ex]
\hline
$Tweets$ & $Proportion$ & $event$ & $slope_p$ & $intercept_p$\\
\hline
\hline

  & & Proposal & 1.81e-11*** & 4.55e-07*** \\
 &$P_{SF}$  & Approval & 1.49e-02* & 5.54e-03**  \\
  & & Enforcement & 5.00e-09*** & 6.39e-10***  \\
\cline{2-5}

 & & Proposal & 1.75e-02* & 1.71e-01  \\
 &$P_{positive}$  & Approval & 1.04e-02* & 3.81e-03** \\
 & & Enforcement & 2.52e-07*** & 7.68e-08*** \\
\cline{2-5}

E-cigarette & & Proposal & 2.28e-02* & 8.54e-05***  \\
&$P_{neutral}$  & Approval & 3.58e-01 & 6.96e-01  \\
 & & Enforcement & 7.99e-09*** & 4.05e-10***   \\
\cline{2-5}

 & & Proposal & 9.64e-07*** & 5.07e-08***  \\
&$P_{negative}$  & Approval & 5.90e-03** & 6.85e-06***  \\
 & & Enforcement & 2.74e-02* & 4.91e-03**   \\
\hline
\hline

% Tobacco --------
 & & Proposal & 8.18e-05*** & 5.67e-01 \\
&$P_{SF}$  & Approval & 2.24e-03** & 5.49e-04*** \\
 & & Enforcement & 2.77e-04*** & 1.36e-04***  \\
\cline{2-5}

 & & Proposal & 1.26e-14*** & 3.24e-17***   \\
&$P_{positive}$  & Approval & 2.07e-13*** & 8.42e-13***  \\
 & & Enforcement &  1.17e-01 & 4.72e-02*  \\
\cline{2-5}

Tobacco & & Proposal & 6.16e-11*** &  2.05e-14***  \\
&$P_{neutral}$  & Approval & 2.96e-10*** & 6.19e-10***  \\
 & & Enforcement & 1.31e-02* & 2.23e-03**  \\
\cline{2-5}

 & & Proposal & 6.90e-08*** & 8.86e-08*** \\
&$P_{negative}$  & Approval & 2.03e-02* & 1.80e-01 \\
 & & Enforcement & 3.16e-02* & 5.52e-03**  \\
\hline
\hline

% Flavored Tobacco --------
 & & Proposal & 2.42e-04*** & 1.59e-01 \\
&$P_{tobacco}$  & Approval & 5.63e-02. & 2.72e-03** \\
 & & Enforcement & 4.23e-06*** & 2.14e-06***  \\
\cline{2-5}

 & & Proposal & 6.18e-12*** & 2.05e-11***   \\
&$P_{positive}$  & Approval & 4.87e-11*** & 1.79e-11***  \\
 & & Enforcement &  1.84e-07*** & 6.38e-08***  \\
\cline{2-5}

Flavored Tobacco & & Proposal & 1.40e-02* &  1.74e-02*  \\
&$P_{neutral}$  & Approval & 1.06e-04*** & 1.03e-08***  \\
 & & Enforcement & 2.64e-14*** & 1.83e-15***  \\
\cline{2-5}

 & & Proposal & 2.44e-12*** & 4.33e-12*** \\
&$P_{negative}$  & Approval & 6.23e-07*** & 1.11e-02* \\
 & & Enforcement & 1.13e-03** & 2.94e-04***  \\
\hline
% \addlinespace[0.5ex]

\multicolumn{4}{l}{
\textsuperscript{***}$p<0.001$, 
\textsuperscript{**}$p<0.01$, 
  \textsuperscript{*}$p<0.05$, 
  . $p<0.1$}
\end{tabular}
}
\end{table}

\subsection{Experiment One Results}
\paragraph{Descriptive Results}
Table \ref{table:summary_table_exp_1} summarizes the average number of Tweets, users, and properties of those users by month during each period of the policy implementation. The total volume of Tweets and users did not change dramatically between different phases of the policy implementation. Visually, the proportion of all e-cigarette Tweets in SF decreases in the first interval, and increases after that (the fourth interval has too few data points to give robust results so we do not include it in any further analysis or discussion) (Figure \ref{fig:exp1_ecig_tobacco}). A similar decrease occurs in the first interval for tobacco Tweets. Flavored tobacco Tweets, in contrast, increase in the first interval and decrease in the second. Each of the changes occurs immediately rather than in a delayed or gradual manner, indicating that a linear regression analysis is suitable. Regression results show consistently that negative and positive trends have significantly different slopes ($p<$0.05 for interaction terms in all three comparisons, for e-cig, tobacco and flavored tobacco Tweets, accounting for other offline and online parameters such as population and proportion of Tweets compared to different denominators (regression results for each outcome not shown, due to space constraints).

Results from the t-test comparing proportions of Tweets before and after the proposal, approval and enforcement of the ban are in Table \ref{table:t-test}. 
Overall, for e-cigarette Tweets in San Francisco, the slope and intercept are significantly different between all time periods. The proportion of e-cigarette Tweets that are positive and negative have significant slope differences for all events.  These differences were not reflected by neutral Tweets, which had no significant slope and intercept differences at all events for e-cigarette Tweets. For the tobacco Tweets, positive, neutral and negative Tweets had significant slope and intercept differences for all events except enforcement (not significant different slope for positive), and approval did not have a significantly different intercept for negative. Flavored tobacco Tweets showed significant differences in slope and intercept for all sentiment groups, but not for overall trends at proposal and approval. For all Tweet categories, the notable difference was decrease in positive Tweets after the proposal was approved, and increase in negative Tweets.

\subsection{Content Analysis}
\paragraph{Qualitative Results From Twitter}
Given these findings about significant changes in Tweeting patterns on e-cigarettes, tobacco and flavored tobacco during and after the flavor ban, we examined Tweets near these time periods to assess what was being discussed. When the proposal occurred (06/2017), there was a significant increase in tobacco and e-cigarette Tweets in San Francisco. Flavored tobacco Tweets were already consistently high, which may explain why there was no similar change at this time. Example Tweets include: \emph{``The People's Socialist Republic of San Francisco has banned flavored eliquid. \#vape \#vapor \#ecig''} (negative sentiment). Notably, we found that the increase in Tweets classified as positive (on e-cigarettes and flavored tobacco after approval) were generally positive on tobacco products (Upon proposal 06/2017; \emph{``SF is voting on whether to ban flavored tobacco, including menthol cigarettes. Wow.''} (labelled positive)). Also, at approval (06/2018); \emph{``I feel like all these anti-vaping ads are having an opposite effect on teens: `wait, there is a bubble gum flavor vape?!'''} and at enforcement
\emph{``Tyr's getting the vapers. Is it February yet?''} (both labelled positive on e-cigarettes).

In contrast, the increasing neutral/negative Tweets (after approval) were were mostly negative about the ban itself or could also include some sarcasm (Upon approval; 06/2018 \emph{``@peterhartlaub I am going to set up an illicit flavored tobacco stand on the platform at the Daly City Bart station. \#misguideddisruption''} (neutral), \emph{``So there’s gonna be a tobacco ban in the city of San Francisco.... sounds like the 1920’s..''} (negative)). Also post approval, enforcement and many months after, saw continued Tweets classified as negative that were negative on the ban; \emph{``It's hard to quit vaping when all your friends vape...''} (negative)), \emph{``Most used phrase of 2018 `where’s my juul?' ''} (neutral). %After enforcement (02/2019): \emph{``My like secret is I smoke a menthol cigarette every six months... and now it’s illegal to buy them in California???…''}.

\begin{table}[b!]
%  \tiny
\centering
\resizebox{0.97\columnwidth}{!}{
\begin{tabular}{m{8.6cm}}
\toprule
I've stated this since the beginning, it's not about the flavors, or the packaging, or the kids... It's about control. (business, proposal, Reddit) \\
\midrule
Name an industry that has multiple years of 70+\% growth, made countless good jobs, and improved the health of millions. \#vape \#ecig \#vaping (business, proposal, Tweet)\\
\midrule
% The reason to point out that the hypocrisy is to show that it isn't about the children. That is what they would have you believe, and have you fight that battle based on some sort of logical "for the kids!" stance. When its actually economic, and about entrenched powers that do not want to allow any give in the fight against ``smoking''. (negative emotion, proposal, Reddit)\\
% \midrule
The last thing you ever want to hear, "I'm from the government and I'm here to help". People are unbelievably stupid these days. (negative emotion, approval, Reddit)\\
\midrule
The FDA is trying to kill vaping because it’s going to do a better job reducing lung cancer than they ever have, an…(negative emotion, approval phase, Tweet)\\
\midrule
If the city of San Francisco wants people to stop juuling so badly they should’ve just banned all flavors except crème brûlée. Big power move. (negative emotion, enforcement, Tweet)\\
\bottomrule

\end{tabular}
} 
\caption{Text of sample Reddit comments and Twitter posts during phases of the San Francisco flavor ban implementation (EMPATH topic, phase, source).}
\label{tab:reddit_post}
\end{table}

\paragraph{Reddit Data and Concordance in Topics}
While temporal confounders can be limited using a time series design, other challenges with using online data as a measurement tool persist. Denominator issues are a challenge in observational data and specifically social media data, for which the contributing population is not representative of the general population \cite{chunara2017denominator}. Accordingly, in order to corroborate and better understand the content of Tweets included at the relevant time periods, and potential reasons for any shifts in sentiment, we sourced data from Reddit, which allows for more nuanced discussion based on its format such as no limits on post length. Reddit is a social news website where registered members can submit content to the site such as links, text posts, and images. Posts are organized by subject into user-created boards , and then other users can then comment on posts, and respond back in a conversation tree of comments. Data from Reddit has been used in numerous social media studies, including to learn about self-disclosure of health related topics including mental health \cite{de2014mental} and conspiracies following dramatic events \cite{samory2018conspiracies}.

In order to find sample Reddit posts (and comments) related to the San Francisco policy change, we first identified related subreddits by using ban related keywords (``San Francisco'' AND Ban, sb140, ``San Francisco'' AND cigarette, ``San Francisco''). Subreddits with the largest numbers of users that contained these keywords, related to tobacco or vaping were: electronic\_cigarette (188K members), ecigclassfields (21.2K members), vaporents (128K members) and Vaping (109K members). We therefore selected the subreddit with the largest number of members. We used the Pushshift Reddit API (https://api.pushshift.io) to collect posts, comments and associated metadata from the electronic\_cigarette subreddit, and then to identify related Tweets, filtered those by any San Francisco or ban keywords (ban, san francisco, sf, sfo, san fran). We remark that this is a very small keyword list compared to the Twitter list, as we specifically want to understand themes of discussion related to the ban; while we can filter using location and time from Twitter, the Reddit subreddits are much more general so we need to filter posts more specifically (filtering by date also returns too many unrelated posts). In order to examine the detailed discussion on the San Francisco ban, we manually checked the title and textual content of each post in the months of the ban proposal, approval or enforcement to make sure the content is relevant, and then crawled all of the comments per post. The dataset contained 16 posts and 607 comments in the proposal month, 10 posts and 577 comments in the approval month, and 8 posts and 110 comments in the enforcement month. Topic analysis was then performed using Empath \cite{DBLP:journals/corr/FastCB16}, on both the Reddit data and filtered Twitter data from the same months. EMPATH is a tool that can generate and validate new lexical categories on demand from a small set of seed terms and capture aspects of affective expression, linguistic style, behavior, and psychological state of individuals from content shared on social media by deep learning a neural embedding across more than 1.8 billion words. EMPATH’s performance has been found to be similar to LIWC (considered a gold standard for lexical analysis), is freely available, and also provides a broader set of categories to choose from \cite{relia2019race}. 

In order to assess similarity or difference with topics in Twitter, we investigated the top overlapping topics overlapping topics between the identified Reddit comments, and filtered Twitter posts from San Francisco during the same months. The top two common topics between the two data sources, identified using EMPATH were negative emotion and business. Example posts in our dataset are given in Table \ref{tab:reddit_post}. We resoundingly found that comments on Reddit also echoed the focus on bans and government setting policy, it was not even possible to find posts on other tobacco/e-cigarette topics, such as people's use of the products, quitting, or other sentiments.

% Summary Table: experiment 2
\begin{table}[h!]
\centering
\caption{Experiment 2: Summary of average number of users, Tweets, and proportion of Tweets by month in each location before and after the tax.
$P_{tweets}$ is the proportion of e-cigarette Tweets in different states.}
\label{table:summary_table_exp_2}
\resizebox{0.95\columnwidth}{!}{
\begin{tabular}{|c|c|c|c|c|}
% if space available
% \addlinespace[1ex]
\hline
$States$ & $Time Period$ & $F_{tweets}$  &$P_{tweets}$ & $F_{users}$  \\
\hline
\hline

% California Level
\multirow{2}{*}{California} & Before Tax &2823 & 3.42e-02 & 1647 \\
\cline{2-5}
 & After Tax & 1863 & 3.63e-02 & 1278  \\
\hline

\multirow{2}{*}{Oregon} & Before Tax & 264 & 4.73e-02 & 140 \\
\cline{2-5}
 & After Tax & 210 & 5.46e-02 & 127 \\
\hline

\multirow{2}{*}{Nevada} & Before Tax & 413 & 6.01e-02 & 158 \\
\cline{2-5}
 & After Tax & 271 & 5.35e-02 & 153  \\
\hline

\multirow{2}{*}{Arizona} & Before Tax & 446 & 3.86e-02 & 311  \\
\cline{2-5}
 & After Tax & 352 & 5.08e-02 & 266 \\
\hline
\hline

% Kansas Level
\multirow{2}{*}{Kansas} & Before Tax & 76 & 2.29e-02 & 55  \\
\cline{2-5}
 & After Tax & 78 & 3.76e-02 & 59  \\
\hline

\multirow{2}{*}{Nebraska} & Before Tax & 63 & 2.49e-02 & 48  \\
\cline{2-5}
 & After Tax & 67 & 4.29e-02 & 55  \\
\hline

\multirow{2}{*}{Missouri} & Before Tax & 209 & 3.49e-02 & 123  \\
\cline{2-5}
 & After Tax & 173 & 4.22e-02 & 115  \\
\hline

\multirow{2}{*}{Oklahoma} & Before Tax & 135 & 2.52e-02 & 95  \\
\cline{2-5}
 & After Tax & 132 & 4.09e-02 & 100 \\
\hline

\multirow{2}{*}{Colorado} & Before Tax & 210 & 3.63e-02 & 142  \\
\cline{2-5}
 & After Tax & 211 & 5.00e-02 & 148 \\
\hline
\hline

% Delaware Level
\multirow{2}{*}{Delaware} & Before Tax & 12 & 1.85e-02 & 10  \\
\cline{2-5}
 & After Tax & 18 & 3.62e-02 & 13  \\
\hline

\multirow{2}{*}{New Jersey} & Before Tax & 347 & 2.81e-02 & 227  \\
\cline{2-5}
 & After Tax & 374 & 4.22e-02 & 236  \\
\hline

\multirow{2}{*}{Pennsylvania} & Before Tax & 359 & 2.75e-02 & 254  \\
\cline{2-5}
 & After Tax & 474 & 4.93e-02 & 295  \\
\hline

\multirow{2}{*}{Maryland} & Before Tax & 221 & 2.09e-02 & 137  \\
\cline{2-5}
 & After Tax & 229 & 3.06e-02 & 173 \\
\hline

\multirow{2}{*}{Washington D.C.} & Before Tax & 82 & 1.91e-02 & 58  \\
\cline{2-5}
 & After Tax & 104 & 2.92e-02 & 66  \\
\hline
\hline

% New Jersey Level
\multirow{2}{*}{New Jersey} & Before Tax & 379 & 3.43e-02 & 237  \\
\cline{2-5}
 & After Tax & 271 & 3.36e-02 & 206  \\
\hline

\multirow{2}{*}{New York} & Before Tax & 1279 & 5.32e-02 & 692  \\
\cline{2-5}
 & After Tax & 981 & 4.16e-02 & 703  \\
\hline

\multirow{2}{*}{Pennsylvania} & Before Tax & 387 & 3.26e-02 & 270  \\
\cline{2-5}
 & After Tax & 502 & 5.53e-02 & 278  \\
\hline

\multirow{2}{*}{Connecticut} & Before Tax & 114 & 2.82e-02 & 84  \\
\cline{2-5}
 & After Tax & 123 & 4.12e-02 & 92  \\
\hline

\multirow{2}{*}{Delaware} & Before Tax & 14 & 2.37e-02 & 11  \\
\cline{2-5}
 & After Tax & 16 & 3.69e-02 & 12  \\
\hline

\multirow{2}{*}{Maryland} & Before Tax & 223 & 2.32e-02 & 149  \\
\cline{2-5}
 & After Tax & 230 & 3.30e-02 & 166  \\
\hline

\end{tabular}
}
\end{table}

\section{Experiment Two: E-cigarette Tax Policy}
%\background on policy change details
\subsection{Background on the E-cigarette Tax Policy}
Many states have and will in the near term enact e-cigarette tax policies. From 2017 to 2018 (overlapping our available Twitter data pull), there were four states which implemented e-cigarette tax policies: California, Kansas, Delaware and New Jersey. In California, the e-cigarette tax policy was effective in 4/1/2017, while Kansas has its e-cigarette tax policy enacted in 7/1/2017, Delaware in 1/1/2018, and New Jersey in 9/29/2018. The exact taxation approach and amounts vary; some by percentage of the wholesale value, while others tax per mL of e-liquid \cite{taxfound}.

\subsection{Natural Experiment Design}
%\bayesian argument
A natural experiment is a condition within the observed dataset which approximates a randomized experiment. An exogenous change, such as a user interface change in a social media system, can be used to create this approximated experiment \cite{oktay2010causal}. Here, we use an external change (tax enactment) for the experiment. It should be noted that considering the before and after for a tax is more appropriate than temporal trends as in experiment one, as a ban takes time to get implemented as different retail locations may get up to speed. However, a tax can be implemented at one point in time. This analysis is also appropriate because there are few other events, exogenous or endogenous to social media data that could be hypothesized to change e-cigarette Tweets by location, thus mitigating spatial, unmeasured confounders. In other words, the probability of tweeting about e-cigarettes compared to the probability of tweeting in state 1 versus 2, will not be affected by any other variables, $P(E_{S1}|H_{S1}) \propto P(E_{S2}|EH_{S2})$.

\subsection{Before and After Comparisons}
As described above, the ratio of e-cigarette taxes between places is of relevance here. Further, we are concerned with how e-cigarette discussion online changes after tax policies (in contrast to absolute levels of Tweeting). Thus, we are comparing ratios of e-cigarette discussion (after to before a tax is enacted), between places with and without a tax. E-cigarette Tweets were filtered using the same procedure as in the interrupted time series experiment. We select geographically adjacent states both to decrease other confounders (geographic trends or preferences) and to examine sentiment in places that may be contextually aware of the tax or places where there could be a consequence. Results are compared using a $\chi^2$ analysis to test the null hypothesis that there is no difference in the after to before ratio of e-cigarette Tweets in places with an e-cigarette tax compared to those with no tax. We tested these relationships separately for positive, neutral and negative sentiment e-cigarette Tweets. Given that adjacent states may also still have underlying contextual differences (e.g. different populations, tweeting populations, etc.) in order to control for these we also use a regression analysis to assess significance of each of these factors. These variables are included as covariates in a regression along with frequency of e-cigarette Tweets before the tax, where the outcome is frequency of e-cigarettes Tweets after the tax.

\begin{table*}[h!]
\centering
\caption{Frequency of positive/negative/neutral e-cigarette Tweets after compared to before tax, and $\chi^2$ and $p$ value for differences between adjacent states.}
\label{table:chi_square_sentiment_frequency}
\resizebox{0.95\textwidth}{!}{
\begin{tabular}{|c|c|c|l|c|c|l|S[table-format=1.2, table-column-width=14mm]|c|l|c|}
% if space available
% \addlinespace[1ex]
\hline
$State$ & $State_B$ & $F_{positive, B}$ & $\chi^2$ & $p$ & $F_{neutral, B}$ & $\chi^2$ & $p$ &  $F_{negative, B}$ & $\chi^2$ & $p$ \\
\hline
\hline

% Maybe remove Nevada
\multirow{3}{*}{California} & Oregon & 0.67 & 0.06 & 0.81 & 0.84 & 4.13 & 0.04* & 1.00 & 0.46 & 0.50  \\
\cline{2-11}
 & Nevada & 0.64 & 0.00 & 0.99 & 0.58 & 0.18 & 0.67 & 1.00 & 0.56 & 0.45 \\
\cline{2-11}
 & Arizona & 0.73 & 0.82 & 0.37 & 0.75 & 2.64 & 0.10 & 0.99 & 0.98 & 0.32  \\
\hline
\hline

% Maybe remove Missouri and Colorado
\multirow{4}{*}{Kansas} & Nebraska & 1.00 & 0.00 & 0.99 & 0.96 & 0.02 & 0.90 & 1.55 & 0.00 & 0.95  \\
\cline{2-11}
 & Missouri & 0.81 & 0.05 & 0.82 & 0.76 & 0.43 & 0.51 & 1.14 & 0.05 & 0.83 \\
\cline{2-11}
 & Oklahoma & 0.83 & 0.02 & 0.89 & 0.98 & 0.00 & 0.98 & 1.23 & 0.00 & 0.97  \\
\cline{2-11}
 & Colorado & 0.93 & 0.02 & 0.90 & 0.95 & 0.01 & 0.91 & 1.29 & 0.00 & 0.96 \\
\cline{2-11}
\hline
\hline

% Maybe remove NJ and Penn
\multirow{4}{*}{Delaware}  & New Jersey & 0.98 & 0.00 & 0.99 & 1.11 & 0.12 & 0.73 & 1.15 & 0.05 & 0.83  \\
\cline{2-11}
 & Pennsylvania & 1.07 & 0.01 & 0.91 & 1.45 & 0.01 & 0.91 & 1.43 & 0.00 & 0.97  \\
\cline{2-11}
 & Maryland & 0.92 & 0.01 & 0.92 & 1.00 & 0.27 & 0.60 & 1.33 & 0.00 & 0.97  \\
\cline{2-11}
 & Washington D.C. & 1.37 & 0.05 & 0.82 & 1.15 & 0.06 & 0.81 & 1.40 & 0.00 & 0.95  
\\
\hline
\hline

\multirow{5}{*}{New Jersey} & New York & 0.65 & 0.58 & 0.45 & 0.72 & 1.54 & 0.22 & 1.19 & 0.89 & 0.35  \\
\cline{2-11}
 & Pennsylvania & 0.99 & 2.03 & 0.16 & 1.53 & 37.50 & 0.00*** & 1.21 & 0.67 & 0.41  \\
\cline{2-11}
 & Connecticut & 1.08 & 1.58 & 0.21 & 1.00 & 4.44 & 0.04* & 1.30 & 0.56 & 0.46  \\
\cline{2-11}
 & Delaware & 1.00 & 0.00 & 0.96 & 0.86 & 0.10 & 0.75 & 1.67 & 0.13 & 0.72 \\
\cline{2-11}
 & Maryland & 0.93 & 0.89 & 0.35 & 0.99 & 6.72 & 0.01* & 1.30 & 0.95 & 0.33  \\
\hline
% \bottomrule
% \addlinespace[0.5ex]

\multicolumn{4}{l}{
\textsuperscript{***}$p<0.001$, 
\textsuperscript{**}$p<0.01$, 
  \textsuperscript{*}$p<0.05$, 
  . $p<0.1$}
\end{tabular}
}
\end{table*}

\subsection{Experiment Two Results}
Table \ref{table:summary_table_exp_2} gives the average number of Tweets, users, and properties of those users by month during each period of the policy implementation. Table \ref{table:chi_square_sentiment_frequency} shows the $\chi^2$ results of the after-to-before e-cigarette Tweeting differences between states which had an e-cigarette tax enacted and adjacent states. $State$ indicates a state which had a tax policy on e-cigarettes enacted at some time during the time period considered, $State_B$ indicates adjacent states of $State$. $F_{sentiment, B}$ is the average number of $sentiment$ Tweets in $State_B$ in one month after, compared to before the tax in $State$, with $sentiment$ as each of: positive, neutral and negative. 

We consistently found that the ratio of negative e-cigarette Tweets after to before the tax was greater in states adjacent to places with a tax (except in Arizona which trailed closely at a ratio of 0.99). On the other hand, positive Tweets generally decreased. For states with the tax, California saw an increase in positive, neutral and negative Tweets (940/600, 1420/881, 461/381). Kansas and Delaware both had a small number of tweets, but saw the increase in negative and decrease in positive (Kansas: 26/24, 36/34, 13/18; Delaware: 4/5 \& 5/8 \& 2/4) and New Jersey saw a decrease in positive Tweets with no significant change in negative Tweets though it should be noted that overall tweeting went down in New Jersey in this period so the proportion actually increased (117/87, 191/116, 70/68). Results of the chi-square tests show a significant difference in the frequency of neutral Tweets in a few locations (Oregon compared to California, and Pennylsvania, Connecticut and Maryland compared to New Jersey). As the neutral Tweets are all Tweets not positive or negative, this may indicate a decrease in overall polarization. It should be noted that sample size of e-cigarette related Tweets per month in Kansas and Delaware was very low. Kansas, and Delaware also the lowest tax per milliliter rate \cite{taxfound}. Therefore, results in those states and adjacent ones are not strong enough to interpret. 

Once we controlled for state-level covariates, the frequency before was significant for predicting frequency after (positive tweets) (Table \ref{tab:regression}). For negative Tweets, a state having enacted a tax was significant predictor for Tweets after. Additionally, the average number of Tweets posted in one month ($F_{tweets})$ as well as $F_{tweetPosters}$, the average number of posters in Twitter in one month, were significant.

\subsubsection{Qualitative Results} Though the frequency of e-cigarette Tweets by month was quite similar before and after the tax, qualitatively we did find increases in Tweets in the months around the tax increase. These often were attributed to an increase in discussion about the tax or cost. For example, e-cigarette related Tweets in Pennsylvania and New Jersey increased specifically in the months leading up to and after the Delaware ban (01/2018). We examined these Tweets manually and found that many were cost-related and commented on the tax. As well, Tweets in Nevada increased upon the ban in California (04/2017). A substantial amount of these discussed cost and/or the tax. Some examples Tweets from the two months after the ban are shared below.

\noindent \underline{Pennsylvania}: \emph{``Do I really wanna waste \$40 on a juul?''}, \emph{``Yeo who’s selling juuls for the lowski I lost mine lastnight and I’m not paying 60 bucks for another one}, \emph{can someone venmo me 50 dollars i wanna buy a juul''}. \underline{New Jersey}: \emph{``Juuls are about to be cancelled bc big tobacco is raising the prices so marbolo lights I will be back for you soon''}. \underline{Nevada}: \emph{``BUY 1 GET 1 FREE on all 60ml's and 120ml's''}.

\begin{table*}[!b]
% \captionsetup{font=small} % why??
\centering
% \small
\resizebox{\textwidth}{!}{
\begin{tabular}{l l l l l l l l l l}
\toprule
% \hline
\textbf{Variable} & $\beta_{positive}$ & std. err & $p$& $\beta_{neutral}$ & std. err & $p$& $\beta_{negative}$ & std. err & $p$\\
% \hline
\midrule

$Constant$ & -19.14 & 8.31 & 4.20e-02* & -37.9572 & 39.68 &  3.59e-01 &  19.81 & 10.17 &  7.70e-02.  \\
$Value_{before}$ & 1.35 & 2.52e-01 & 0.00*** & -2.06 & 1.23 &  1.23e-01 &  8.53e-01 &  6.41e-01 &  2.10e-01\\
$Tax$ & 2.51 & 3.01  & 4.23e-01 & 16.47 & 14.74 &  2.88e-01 &  -9.56 &  3.53 &  2.00e-02*\\
$Population$ & 9.59e-07  & 1.14e-06 & 4.20e-01 & 1.57e-05  & 5.65e-06 &  1.80e-02* &  2.31e-06 &  1.53e-06 &  1.60e-01   \\
$F_{tweets}$ & 2.53e-05  & 9.44e-06  & 2.10e-02* & -1.98e-06 & 4.50e-05 & 9.66e-01 &  -6.01e-05  & 1.13e-05 &  0.00***   \\
$F_{users}$ & 7.00e-04 & 0.00  & 6.00e-03** & 5.00e-04&  1.00e-03 &  5.83e-01 & 9.00e-04 &  0.00 &  4.00e-03** \\
$F_{ecigTweets}$ & -5.97e-01 & 1.30e-01  & 1.00e-03** & 1.04&  6.93e-01 &  1.60e-01 &  3.10e-02 &  5.90e-02 &  6.12e-01 \\
$F_{ecig\_users}$ & 3.25e-01 & 9.80e-02 & 7.00e-03** & 4.39e-02&  3.89e-01 &  9.12e-01 & -4.70e-03 &  2.06e-01 &  9.82e-01\\
$P_{ecigTweetsstate}$ & 558.39 & 257.38  & 5.30e-02. & 278.14 &  1237.91 &  8.26e-01 &  -547.03 & 301.54 &  9.70e-02.\\
% \hline
% \hline

\midrule
\bottomrule
% \addlinespace[1ex]
\multicolumn{4}{l}{
\textsuperscript{***}$p<0.001$, 
\textsuperscript{**}$p<0.01$, 
  \textsuperscript{*}$p<0.05$, 
  . $p<0.1$}
\end{tabular}
}

\caption{Regression results for frequency of e-cigarette Tweets in different sentiment before and after the tax date. $F_{tweets}$ is avg number of Tweets posted per month, $F_{users}$ avg number of posters in Twitter in one month, $F_{ecigTweets}$ indicates the average number of e-cigarette Tweets per month, $F_{ecig\_users}$ is avg number of user that post e-cigarette Tweets per month, and $P_{ecigTweetsstate}$ is proportion of e-cigarette Tweets in all Tweets from the same state.}
\label{tab:regression}
\end{table*}

\section{Discussion and Conclusion}
Our study focused on assessing the response on social media to offline policy changes, during narrow time periods and in specific locations. We used quasi-experimental designs to mitigate temporal and spatial confounders. We examined the content of resulting Tweets and compared to other contemporaneous online data (from Reddit) to corroborate findings on discussion with respect to the policy changes. Overall, it was clear that negative sentiment and discussion about the policies dominated discussion, and increased directly after policy implementation. We did not find evidence for increased polarization online in places adjacent to where a tax was implemented, however, within the same state, once controlled for state-level online data and population covariates, a state having enacted a tax was significant predictor for negative Tweets after the tax. The findings regarding a focus on policy and anti-government related discussion differs from other work which found more discussion related to the \textit{content} of policies (e.g. for abortion). However, these are different types of policies; abortion has a longer history and ideology behind it, while tobacco regulations are rapidly changing (and harder to measure via surveys for this reason). Elaboration on findings regarding what was discussed online in response to policy changes are discussed below. Given the focus on policy-related discussion, and that Twitter users are approximately one-fifth of the United States population, and how the people included in the data here compare to the broader population is not well understood, we also distill specific utility of high-resolution and location-specific unobtrusive online data in response to policy changes at the end of this section \cite{perrin2019share}.

In the first experiment we used an interrupted time series design and segmented regression to reduce threats to validity such as historical Tweet trends. The proportion of flavored tobacco Tweets in SF increased significantly after the flavor ban proposal. Conversely, the proportion of Tweets on e-cigarettes increased significantly after proposal approval. There also was a significant difference in tobacco, e-cigarette and flavored tobacco sentiment in San Francisco on Twitter at events during the regulatory period (regulation proposal, approval and enforcement), particularly more polarized Tweets, which was driven by negative sentiment. Within these periods, sentiment towards flavored tobacco, tobacco and e-cigarettes also showed differences; for example after proposal approval, positive Tweets on all categories decreased, while negative Tweets increased. Examining these changes qualitatively, we found that initially, after the flavored tobacco proposal, people discussed flavored tobacco and products positively. This shifted to negative discussion on the ban itself after approval, which was corroborated by discussion on Reddit. In other words, we did not find significant evidence that the increase in positive discussion on flavored tobacco after the proposal was about the ban. Notably, the increased volume of e-cigarette Tweets was not flavor-specific, instead a general increase in e-cigarette discussion. 

In the second experiment we use a natural experiment design to assess how enacting of an e-cigarette ban in a state relates to sentiment changes in Twitter while controlling for other unmeasured confounders by location. Comparing the ratio of e-cigarette Tweets (by sentiment) after-to-before the tax in states which had a tax enacted, compared to adjacent states we found a significant difference in this ratio in most of the adjacent states to New Jersey, and in Oregon, in relation to California \ref{table:chi_square_sentiment_frequency}). 
%We found a significant change in polarization after tax enactment (Table \ref{table:chi_square_sentiment_frequency}). 
When controlling for other covariates using a regression analysis to more closely replicate a random experiment, we found that a change in frequency of negative Tweets was predicted by having a tax (controlling for state covariates) was significant. Qualitative examination of Tweets before and after the tax showed increased discussion of cost in states adjacent to those with the tax. 

\subsection{Limitations and Future Work}
We found that the number of geo-located Tweets sourced from the API on specific topics, such as e-cigarettes, became low in certain places (e.g. frequency of e-cigarette Tweets in Kansas and Delware). Ideally for this experiment we would have used proportion of all Tweets that were e-cigarette related, by state, but these low numbers prevented us from doing so. This is an established concern about using geo-located data from the free API \cite{relia2018socio}. However, given that larger samples of Twitter data are available (e.g. from the firehose), this could be easily remedied. Further, although we use quasi-experimental designs to isolate effects in online social media data as best as possible, there are other possible challenges. Indeed, there could be other offline confounders resulting in changes in the data. However, the designs specifically serve to reduce those (confounders at the same time as the events of interest in experiment one and differences between places that would confound the before to after e-cigarette discussions in experiment two).

Along with the quantitative results, qualitative investigation shows that discussion online can indicate important aspects in response to and about policy changes. For example, in experiment one, negativity \emph{about the ban} started to increase, dominating any conversation that was negative on tobacco products. As well, we found discussion about cost increases in states adjacent to those that enacted taxes.; future work developing nuanced topic modeling on policy may help unpack these trends. 

These results can be put into practice in several ways. We examined possible other datasources regarding public response to tobacco and e-cigarette policies. While we could not find any survey asking people's attitude towards an e-cigarette flavor policy (it may be too early for any surveys to have been deployed), there are some national surveys asking the \textit{existence} of tobacco-free policy. For example, the 2016 School Health Policies and Practices Study which is a national survey that assess school health policies and practices at the state, district, school, and classroom levels. This survey has a question on ``Has your district adopted a policy prohibiting cigarette smoking by students?'' \cite{centers2016school}. The 2019 National Survey on Drug Use and Health asks ``At your workplace, is there a written policy about employee use of alcohol or drugs?'' \cite{abuse2014national}. While such national-level representative surveys do not ask about sentiment on specific policies, some more local and targetted surveys have asked people's attitude towards e-cigarette or tobacco policies. Each of the surveys we found reported strong public support for policies; in Hong Kong, Saint Paul Minnesota, and the United States in general, there was very strong public support for new policies that have been implemented in other countries, like plain packaging, banning point-of-sale tobacco displays, and increasing the legal age of purchasing cigarettes. One study in New York City showed dramatically lower support by current smokers \cite{cheung2018report,saintpaul2014,winickoff2016public,farley2015public}. These mixed but generally positive support towards policies, as well as variation by group, in contrast to our findings of negative sentiment around policies and government, help distill the role of social media. As survey methods can also be used to assess public opinion but are often time constrained and populations may not disclose their opinions on them. Thus, information from the online population, available in real-time, can provide a window towards initial responses that may not get captured in data sources such as surveys. While we do not advocate that sentiment from this population should be generalized to the whole population, the striking negative sentiment around policies during rapid shifts should be investigated further, and viewed as a complementary source of information regarding public response. Detailed follow up or recruitment from online locations could help learning in more detail the drivers of sentiment and background of the individuals (both demographic and more complex characteristics). While tobacco and e-cigarettes are a prime exemplar of a rapidly changing policy environment, this use-case can be extended to other settings. In sum, we advocate for using social media data as one signal to understand response to policy changes, and potentially a way to find populations with views that don't get captured on traditional survey mechanisms. Multiple datasets and sources must be aggregated together in order to get a full view of population sentiment.

\section{Appendix}
\subsubsection{E-cigarette keywords:} \underline{Synonyms:} electronic-cigarette, electronic cigarette, electronic cig, e-cig, ecig, e cig, e-cigarette, ecigarette, e cigarette,e cigar, e-juice, ejuice, e juice, e-liquid, eliquid, e liquid, e-smoke, esmoke, e smoke, vape, vaper, vaping, vape-juice, vape-liquid, vapor, vaporizer, boxmod, cloud chaser, cloudchaser, smoke assist, ehookah, e-hookah, e hookah, cherry tip cigarillo, stillblowingsmoke, still blowing smoke,  smoke \& pod. \underline{Brands:} juul, vaporfi, vype pebble, v2 cig, v2cigs, v2 cigs, halocigs, njoy, markten, vuse, tryst, atomizer, cartomizer, south beach smoke, eversmoke, joye510, joye 510, joyetech, logicecig, smartsmoker, mistic, smokestiks, 21st century smoke logic black label, finiti , nicotek, cigirex, ciga \& blu, cig \& blu, logic \& cig, e-swisher, e swisher, eswisher, ezsmoker, ez \& cig, green smoke, cigalectric, xhale o2, xhaleo2, cig2o, green smart living, greensmartliving, swisher blk, grimmgreen, \#njoy, \#fin, \#finiti. \underline{Policies:} sb140, sb 140, sb24, sb 24. \underline{Cessation:} notblowingsmoke, not blowing smoke, tobaccofreekids, notareplacement.

\subsubsection{Flavored e-cigarette keywords:} blueberry, pomegranate, strawberry, orange, grapefruit, kiwi, guave, lemonade, apple, mango, peach, pineapple, raspberry, mint, lemon, grape, watermelon, fruit, citrus, lime, banana, coconut, berry, dragon fruit, melon, cherry, menthol, vanilla, candy, gummy, cotton candy, butterscotch, candy cane, caramel, tart, cheesecake, cinnamon roll, macaron, cream, cookie, cake, coffee, espresso, milk, cracker, mocha, cappuccino, cocoa, flavor, pod, eon pod, cali pod, sea pod, ziip pod.

\subsubsection{SF ban keywords:}
ban, san francisco, sf, sfo, san fran.
% not_flavor_ecig = ['smoky', 'smoked']

\bibliography{policy}
\bibliographystyle{aaai}
\end{document}